\documentclass[conference]{IEEEtran}

\usepackage{graphicx}%
\usepackage{bm}%
\usepackage{amsfonts}
\usepackage{amssymb}
\usepackage{times}
\usepackage{subfigure}
\usepackage{enumitem}
\usepackage{amsmath} 
\usepackage{latexsym}
\usepackage{hhline}
\usepackage{cite}
\usepackage{caption}
\usepackage{url}
\usepackage{booktabs}
\usepackage[linesnumbered,ruled,vlined]{algorithm2e}
\usepackage{multirow} 
\usepackage{xcolor}
\usepackage{epstopdf}
\usepackage{aligned-overset}

\allowdisplaybreaks

\begin{document}
\bstctlcite{bibfbcs:BSTcontrol}
\title{
Learning-Aided Short Code Design for ISAC based on MIMO-OFDM 
}
\author{\IEEEauthorblockN{
Mingcheng Nie\IEEEauthorrefmark{1},
Shuangyang Li\IEEEauthorrefmark{2}, 
Geng Wang\IEEEauthorrefmark{1}, 
Peng Cheng\IEEEauthorrefmark{3}, 
Shenghong Li\IEEEauthorrefmark{4},
Chang Liu\IEEEauthorrefmark{3}, \\
Giuseppe Caire\IEEEauthorrefmark{2}, 
and
Yonghui Li\IEEEauthorrefmark{1}
}
\IEEEauthorblockA{
\IEEEauthorrefmark{1}The University of Sydney, Sydney, Australia\\
\IEEEauthorrefmark{2}Technische Universit{\"a}t Berlin, Berlin, Germany\\
\IEEEauthorrefmark{3}La Trobe University, Australia\\
\IEEEauthorrefmark{4}Data 61, CSIRO, Australia 
}}


\maketitle
\begin{abstract}
This paper proposes a deep learning (DL)-based coded waveform design for integrated sensing and communications (ISAC), enabling flexible trade-offs between communication reliability and ranging accuracy in short-block transmissions. The proposed scheme is built upon a practical multiple-input multiple-output orthogonal frequency-division multiplexing (MIMO-OFDM) architecture, where the communication channel state information and the angles of the static targets are assumed available at the transmitter. A transformer-based transmitter encodes input information bits directly into ISAC transmit waveforms to jointly optimize the bit error rate (BER) performance and the delay modified Cramér-Rao bound (MCRB). A corresponding transformer-based receiver is adopted at the communication side to recover the transmitted information bits. We further examine the learned codewords for communication-oriented and sensing-oriented designs, revealing that a balanced ISAC waveform naturally exhibits an intermediate structure between these two extremes. Numerical results illustrate these codeword structures and demonstrate that the proposed design provides substantial trade-off gains over conventional schemes based on standard channel coding and modulation.

\let\thefootnote\relax\footnotetext{
The work of Shuangyang Li was supported in part by the European Research Council (ERC) under the ERC Starting Grant No. 101220383 (Foundations of Delay Doppler Communications and Sensing, FUNDOCS).   }
\end{abstract}

\section{Introduction}
International Mobile Telecommunications 2030 (IMT-2030) identifies sensing as a core capability of sixth-generation (6G) networks to enable a broad range of emerging applications and use cases~\cite{nie11373535standard}. To realize this vision and meet the stringent sensing-related key performance indicators (KPIs), integrated sensing and communication (ISAC) is expected to play a crucial role by optimizing spectrum utilization while reducing deployment and operational cost. These benefits are primarily enabled by the joint design of sensing and communication waveforms, which facilitates a scalable trade-off between the two functionalities~\cite{liu2022integrated}. However, joint waveform design remains fundamentally challenging because communication and sensing impose different, and often conflicting, signal requirements. For example, communication systems
rely on randomness to convey information efficiently, whereas sensing systems favor deterministic signals to ensure stable sensing performance \cite{xiong2023fundamental}.

To address this challenge, deep learning (DL) has recently received considerable attention. With strong nonlinear representation capability, DL can learn complex mappings from input bits to transmit waveforms while simultaneously accommodating both communication and sensing objectives. Consequently, DL offers the potential to explore solution spaces that are difficult to tackle using conventional approaches due to the complexity of the underlying optimization problems \cite{nie2026neural}. In this context, a growing number of studies have explored DL-based coded waveforms for ISAC. In \cite{kim2024short}, a multilayer perceptron (MLP) was used to design short-block length codes for non-coherent communication detection and coherent radar sensing. The learned codewords exhibit distinct structures: phase shift keying (PSK) signaling appears in sensing-oriented code, whereas on-off keying emerges in communication-oriented code. By adjusting the relative weight of the two objectives, the resulting ISAC codewords interpolate between these limiting cases. A similar design philosophy was extended to orthogonal frequency-division multiplexing (OFDM) systems in \cite{bian2025lisac}, where recurrent neural networks (RNNs) were employed to capture sequential correlation across codewords. To reduce training complexity, a long codeword is partitioned into multiple blocks, each of which is fed into an RNN. However, inter-block dependencies are not fully observed in this training structure. The learned codewords exhibit similar trends, with a PSK-like constellation for sensing-oriented designs and a more dispersed constellation for communication-oriented designs, which carries more information and increases the distance between codewords. Compared to \cite{kim2024short,bian2025lisac}, where sensing loss is defined using the mean square error (MSE) criterion, the authors in \cite{aditya2025channel} instead exploit the autocorrelation properties of the learned codewords and formulate the sensing loss in terms of sidelobe levels. Our study falls within this stream of work but distinguishes itself through different systems, channels, and learning models. In particular, we move beyond single-antenna or simplified OFDM settings to a practical multi-antenna system, which requires joint coding across subcarriers while incorporating heterogeneous inputs, thereby introducing new challenges and insights.

In this paper, we propose a novel DL-based coded waveform design framework for ISAC in multiple-input multiple-output OFDM (MIMO-OFDM) systems. The spatial-frequency degrees of freedom provide a natural basis for joint communication and sensing waveform design.  
Since ISAC beamforming for initial target probing has been well studied, and an omnidirectional beam pattern can be generated while satisfying practical communication requirements \cite{liu2018toward}, 
we focus on the post-probing stage where the target angles are available at the transmitter and delay estimation becomes the primary sensing task. 
We further consider a time-division duplexing (TDD) mode, where channel state information at the transmitter (CSIT) for the communication link is available via channel reciprocity.  
Specifically, a transformer-based transmitter jointly maps the information bits across all subcarriers, communication CSI, and target-angle information into a space-frequency ISAC waveform. Dedicated input projections and joint token processing enable the transmitter to combine these different inputs and capture dependencies across subcarriers. The transmitter is trained using a combined loss based on binary cross-entropy and the delay modified Cram\'er--Rao bound (MCRB), with a weighting factor controlling the communication--sensing trade-off. At the communication receiver, a transformer-based decoder recovers the information bits from received-signal and channel features without requiring the sensing angles. Beyond the transceiver architecture, we analyze the learned codeword structures, showing that communication-oriented designs distribute energy across delay bins, whereas sensing-oriented designs produce impulse-like waveforms with power concentrated near the target directions. Our numerical results show that the balanced ISAC codeword can be interpreted as a structured combination of these two extreme cases. Simulation results demonstrate that the proposed scheme achieves favorable BER--MCRB tradeoffs and outperforms conventional transmission schemes based on standard channel coding and modulation.

\emph{Notation:}
$(\cdot)^{\rm{H}}$, $(\cdot)^{\top}$, and ${\rm Tr}\{\cdot\}$ denote the Hermitian transpose, transpose, and trace operation, respectively; 
${\bf I}_M$ represents the $M\times M$ identity matrix.
$\mathbb{E}[\cdot]$ denotes the expectation operation;
The circularly symmetric complex Gaussian distribution having variance $\sigma^2$ is denoted by $\mathcal{CN}(0,\sigma^2)$. $\mathcal{R}(\cdot)$ stacks the real and imaginary components of the complex entries into a real-valued vector.


\section{System Model}

\begin{figure}
    \centering
    \includegraphics[scale=1]{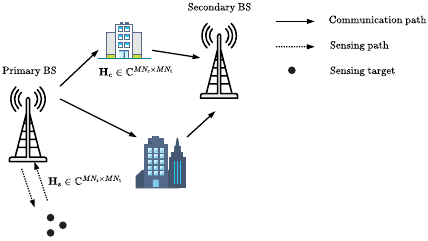}
    \caption{Monostatic ISAC scenario.}
    \label{fig:system_model}
\end{figure}
We consider ISAC transmission in a MIMO-OFDM system with $M$ subcarriers, subcarrier spacing $\Delta f$, and OFDM symbol duration $T$~\cite{nie2025novel}. The system operates under the critical sampling condition such that $\Delta f=\frac{1}{T} $ \cite{nie2026refinement}. As shown in Fig.~\ref{fig:system_model}, each base station (BS) is equipped with $N_{\mathrm{t}}$ transmit antennas and $N_{\mathrm{r}}$ receive antennas. Here, a primary BS transmits an ISAC signal, which experiences a fading channel and is received by a secondary BS for communication purposes, while the echoes reflected from surrounding static sensing targets are simultaneously processed at the primary BS for sensing tasks. 
The communication and sensing channels are mutually independent, and for the $m$-th subcarrier, $0\le m\le M-1$, they are denoted by $\mathbf{H}_{\mathrm{c},m}\in\mathbb{C}^{N_{\mathrm{r}}\times N_{\mathrm{t}}}$ and $\mathbf{H}_{\mathrm{s},m}\in\mathbb{C}^{N_{\mathrm{r}}\times N_{\mathrm{t}}}$, respectively. For the $m$-th subcarrier, a length-$C$ bit sequence $\mathbf{c}_{m}$ is fed into the proposed neural network (NN), which generates a length-$N_{\mathrm{t}}$ ISAC transmit sequence $\mathbf{x}_m$, subject to a per-subcarrier power constraint $\sum\nolimits_{i=1}^{N_{\rm t}}|x_{m,i}|^2=N_{\rm t}$.
After propagation through the communication and sensing channels, the received signals at the $m$-th subcarrier for communication and sensing are denoted by $\mathbf{y}_{\mathrm{c},m}\in\mathbb{C}^{N_{\mathrm{r}}\times 1}$ and $\mathbf{y}_{\mathrm{s},m}\in\mathbb{C}^{N_{\mathrm{r}}\times 1}$, respectively, as
\begin{align}
    \mathbf{y}_{\mathrm{c},m}&=\mathbf{H}_{\mathrm{c},m}\mathbf{x}_{m}+\mathbf{n}_{\mathrm{c},m}, \ \mathbf{y}_{\mathrm{s},m}=\mathbf{H}_{\mathrm{s},m}\mathbf{x}_{m}+\mathbf{n}_{\mathrm{s},m}.
\end{align}
Here, $\mathbf{n}_{\mathrm{c},m}\sim\mathcal{CN}(0,\sigma^{2}_{\mathrm{c}}\mathbf{I}_{N_{\mathrm{r}}})$ and $\mathbf{n}_{\mathrm{s},m}\sim\mathcal{CN}(0,\sigma^{2}_{\mathrm{s}}\mathbf{I}_{N_{\mathrm{r}}})$ denote the complex additive white Gaussian noise (AWGN) at the communication receiver and sensing receiver, respectively. The corresponding channels for the $m$-th subcarrier are given, respectively, as
\begin{align}
    \mathbf{H}_{\mathrm{c},m}&=\sum\nolimits_{p=1}^{P} \tilde{h}_p e^{-j2\pi\tilde{\tau}_p m \Delta f}\mathbf{A}(\tilde{\theta}_p,\tilde{\varphi}_p), \label{H_c_m}
\end{align}
\begin{align}
    \mathbf{H}_{\mathrm{s},m}&=\sum\nolimits_{q=1}^{Q} h_q e^{-j2\pi\tau_q m \Delta f}\mathbf{A}(\theta_q,\theta_q). \label{H_s_m}
\end{align}
Here, $P$ and $Q$ denote the number of resolvable paths in the communication channel and the number of sensing targets, respectively. For the communication channel, $\tilde{h}_p\sim\mathcal{CN}(0,\frac{1}{P})$, $\tilde{\tau}_p$, $\tilde{\theta}_p$, and $\tilde{\varphi}_p$ represent the fading coefficient, delay, angle of arrival (AoA), and angle of departure (AoD) of the $p$-th path, respectively. For the sensing channel, $h_q\sim\mathcal{CN}(0,\frac{1}{Q})$, $\tau_q$, and $\theta_q$ denote the corresponding fading coefficient, delay, and angle associated with the $q$-th target, respectively. The delay resolution depends on the signal bandwidth, such that $\tau=\frac{l}{M\Delta f}$, where $l$ denotes the normalized delay and is not necessarily an integer. The matrix $\mathbf{A}=\mathbf{a}_{\mathrm{r}}(\tilde{\theta}_i)\mathbf{a}_{\mathrm{t}}^{\top}(\tilde{\varphi}_i)$ denotes the array steering matrix. Assuming uniform linear arrays (ULAs), the receive and transmit steering vectors are given by $\mathbf{a}_{\mathrm{r}}(\tilde{\theta}_i)=\sqrt{1/N_{\mathrm{r}}}[e^{-j\pi\frac{(N_{\mathrm{r}}-1)}{2} \sin{\tilde{\theta}_i} }, \dots, e^{j\pi \frac{(N_{\mathrm{r}}-1)}{2}\sin{\tilde{\theta}_i}} ]^{\top}$ and $\mathbf{a}_{\mathrm{t}}(\tilde{\varphi}_i)=\sqrt{1/N_{\mathrm{t}}}[e^{-j\pi\frac{(N_{\mathrm{t}}-1)}{2} \sin{\tilde{\varphi}_i} }, \dots, e^{j\pi \frac{(N_{\mathrm{t}}-1)}{2}\sin{\tilde{\varphi}_i}} ]^{\top}$, respectively. Additionally, the AoA is assumed to be equal to the AoD in the sensing channel due to the monostatic sensing setup. By stacking all received signals across subcarriers, the input-output relation in the frequency domain is expressed as
\begin{align}
    \mathbf{y}_{\rm c} &= \mathbf{H}_{\rm c}\mathbf{x}+\mathbf{n}_{\rm c}, \ \mathbf{y}_{\rm s} = \mathbf{H}_{\rm s}\mathbf{x}+\mathbf{n}_{\rm s}, \label{linear_IOR}
\end{align}
where $ \mathbf{H}_{\rm c}$ and $ \mathbf{H}_{\rm s}$ are block-diagonal matrices, whose subblocks are given in \eqref{H_c_m} and \eqref{H_s_m}, respectively.

The overall procedure consists of two phases. In the first phase, the sensing angles are unknown. Consequently, the transmitted ISAC signal is precoded using a communication-oriented scheme based on the singular value decomposition (SVD) of the communication channel. This design yields a statistically near-omnidirectional beampattern, which is well-suited for initial target probing in sensing tasks when no prior angular information is available~\cite{liu2018toward}. Due to space limitations, the details of this phase are omitted. In this paper, we focus on the second phase. After the initial probing and angle estimation, the target angles are assumed to be exactly known. Under this assumption, the design objective shifts to balancing delay sensing performance and communication error performance, where the precoding stage is implicitly learned within the proposed NN. 

\section{Neural Network Design and Training}

\begin{figure*}[ht!]
    \centering
    \includegraphics[scale=0.62]{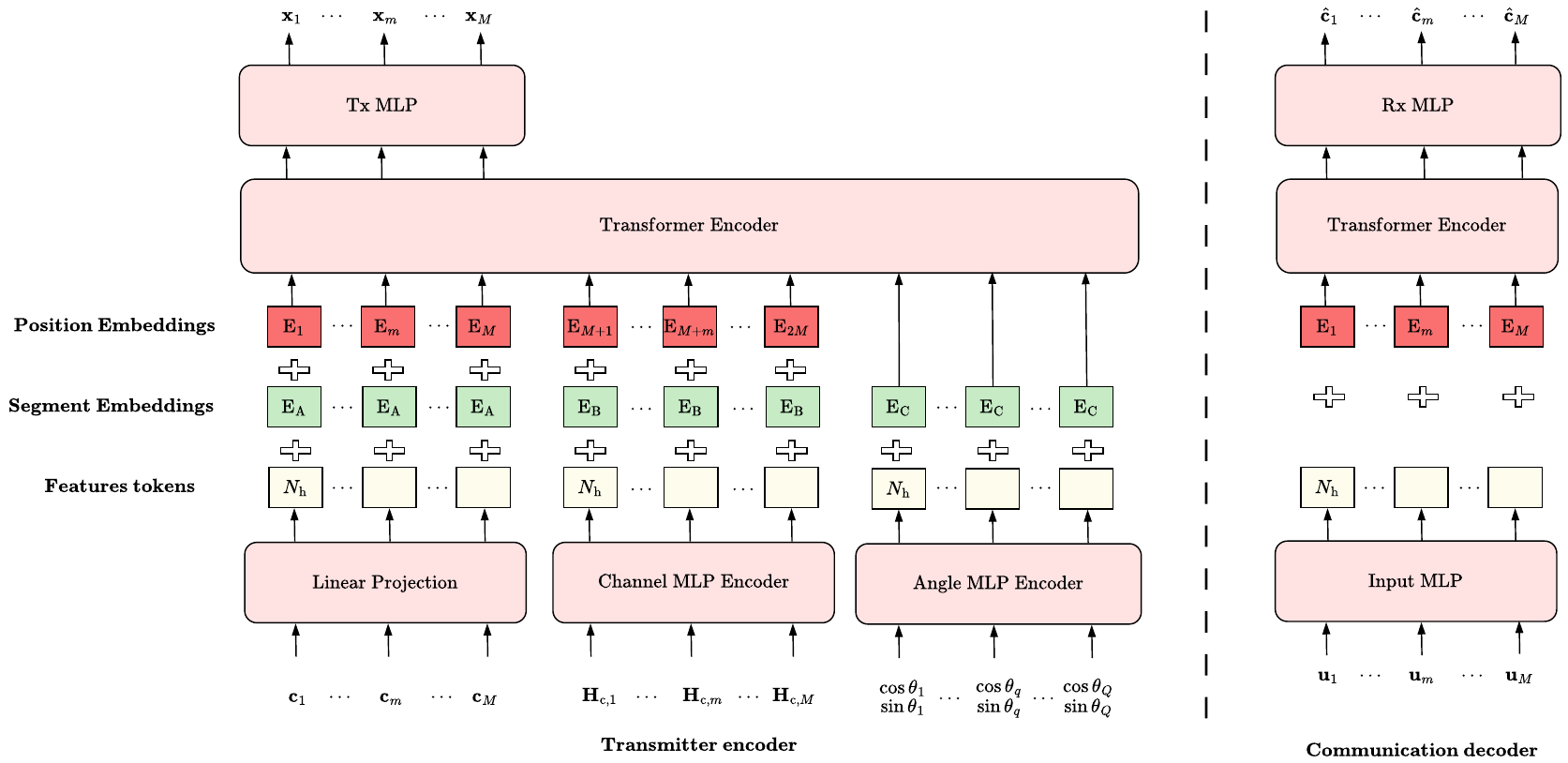}
    \caption{NN model overview. }
    \label{fig:nn_structure}
\end{figure*}
\subsection{Neural Network Architecture}
We consider three types of input for the transmitter encoder: 1) the information bits across all $M$ subcarriers $\mathbf{c}\in \{0,1\}^{M\times C}$;\footnote{Note that the dimension here is expressed in NN tensor form, and differs from the linear model in \eqref{linear_IOR}. However, both formulations represent exactly the same variable.} 2) the communication channel matrices for all subcarriers $\mathbf{H}_{\rm c}\in\mathbb{R}^{M\times 2 N_{\rm r} N_{\rm t}}$; and 3) the angle of sensing targets $\boldsymbol{\theta}\in\mathbb{R}^{Q\times 2}$. The objective of the transmitter NN is to learn the space-frequency ISAC transmit signal $\mathbf{x}\in\mathbb{C}^{M\times N_{\rm t}}$ from these inputs, i.e.,
\begin{align}
     \mathbf{x} \sim f_{\psi} (\mathbf{c}, \mathbf{H}_{\rm c}, \boldsymbol{\theta}),
\end{align} 
where $f_{\psi}(\cdot)$ denotes the transmitter encoder parameterized by $\psi$. Since the inputs are inherently heterogeneous in both data types and dimensionality, efficient feature extraction and fusion for these high-dimensional multi-source inputs is nontrivial. To address these challenges, we employ three dedicated projection modules to map each input into a feature token with dimension $N_{\rm h}$ in a shared latent space:
\subsubsection{Projection for Information Bits $\mathbf{c}$} A linear projection is applied to each information-bit sequence $\mathbf{c}_{m}$, yielding a feature token with dimension $N_{\rm h}$. Unlike \cite{bian2025lisac}, which processes individual bits, we operate on bit sequences such that each feature token represents a length-$C$ bit block. This design reduces the number of feature tokens for the information bits from $MC$ to $M$, thereby improving computational efficiency. Note that the correlations across blocks are preserved by feeding all tokens into the transformer~\cite{vaswani2017attention}. 
\subsubsection{Projection for Communication Channel $\mathbf{H}_{\rm c}$}
A similar approach is applied to extract features from the communication channel matrices $\mathbf{H}_{{\rm c},m}$, yielding $M$ channel tokens with dimension $N_{\rm h}$. Note that the complex-valued channel entries are first decomposed into real and imaginary components before processing.
\subsubsection{Projection for Sensing Angles $\boldsymbol{\theta}$}
Each sensing angle is encoded as $[\sin(\theta_{q}), \cos(\theta_{q})]$ rather than raw angle values. This circular representation avoids discontinuities caused by angle periodicity and provides a smooth feature space for learning. Moreover, it is consistent with the array-response model $\mathbf{a}(\theta_q)$, which depends on $\sin(\theta_q)$, thereby enabling more effective extraction of angle-dependent sensing features.

Segment embeddings are then introduced to distinguish tokens from different input sources by $\mathrm{E}_{\mathrm{A}}$, $\mathrm{E}_{\mathrm{B}}$, and $ \mathrm{E}_{\mathrm{C}}$ \cite{devlin2019bert}, while positional embeddings are incorporated to preserve the structural information of tokens by $\mathrm{E}_{\mathrm{1}}, \dots, \mathrm{E}_{\mathrm{2}M} $. No positional embeddings are introduced for the angle tokens, as they represent an unordered set of physical paths rather than a sequence with intrinsic order. The resulting token sequence is then processed by a transformer encoder for joint feature modeling \cite{vaswani2017attention}, as illustrated in Fig.~\ref{fig:nn_structure}. After feature fusion, only the first $M$ output tokens of the transformer are selected to generate the transmission sequence via an MLP. This is because these tokens correspond to the $M$ information bit tokens, whereas the channel and angle tokens serve as auxiliary tokens to enrich the learned representations rather than being directly mapped to the transmit waveform. A power normalization is imposed such that $\mathbf{x}_{m}^{\rm H}\mathbf{x}_{m} = N_{\rm t}$.

At the communication receiver, the maximum likelihood (ML) receiver is often adopted to achieve optimal performance at the cost of complexity. However, since the sensing angles are typically unavailable at the communication receiver, the possible $\mathbf{x}$ for all information bits cannot be computed, rendering ML receiver infeasible. Therefore, we instead employ an NN-based decoder to recover the transmitted information bits using the received signals and communication channel information. Specifically, the receiver first computes the matched-filter output and channel Gram matrix for each subcarrier \(m\) as
\begin{equation}
    \mathbf z_m
    \triangleq
    \mathbf H_{{\rm c},m}^{\rm H}\mathbf y_{{\rm c},m},
    \quad
    \mathbf g_m
    \triangleq
    \operatorname{vec}(\mathbf H_{{\rm c},m}^{\rm H}\mathbf H_{{\rm c},m}),
\end{equation}
where \(\mathbf z_m\) represents the channel-matched observation, and \(\mathbf g_m\) characterizes the effective channel gains and correlations. These quantities are combined with the channel matrix and noise variance to construct the input feature vector
\begin{align}
    &\mathbf u_m =
    \big[
        \operatorname{vec}(\mathcal{R}(\mathbf H_{{\rm c},m}))^{\top},
        \mathcal{R}(\mathbf z_m)^{\top},
        \mathcal{R}(\mathbf g_m)^{\top},
        \sigma_{{\rm c},m}^{2}
    \big]^{\top}.
\end{align}
An MLP then maps \(\mathbf u_m\) to an
\(N_{\rm h}\)-dimensional feature token, which is then combined with a learnable positional embedding. The resulting \(M\) tokens are fed into a transformer encoder to jointly process the information across all subcarriers. Finally, each token is passed through an output MLP to generate \(C\) bit logits. A sigmoid function maps these logits to soft bit probabilities \(\mathbf p_m\in(0,1)^C\), which are used to compute the training loss. During inference, the hard decision is \(\hat c_{m,k}=1\) if \(p_{m,k}\ge 1/2\), and \(0\) otherwise, for \(k=1,\dots,C\). The recovered bit sequences across all subcarriers are stacked as $\hat{\mathbf{c}}=[\hat{\mathbf{c}}_{1}^{\top},\dots,\hat{\mathbf{c}}_{M}^{\top}]^{\top}$.

\subsection{Design of Performance Metric}

The communication performance is evaluated in terms of the BER using the hard decisions $\hat c_i$. During training, the NN parameters are optimized using the binary cross-entropy (BCE) loss evaluated on the soft probabilities $p_i$, obtained by stacking the vectors $\mathbf p_m$ across all subcarriers:
\begin{align}
    \mathcal{L}_{\mathrm{c}}=-\frac{1}{MC}\sum_{i=1}^{MC}\left[ c_{i}\log{p_{i}} + (1-{c}_{i})\log{(1-p_{i})}\right].\label{comm_loss}
\end{align}

For sensing, the delay estimation performance is quantified using the Cram\'er--Rao bound (CRB), which provides a fundamental lower bound on the estimation error covariance of any unbiased estimator. Let $\mathbf{l}=[l_1,\dots,l_Q]^{\top}$ denote the delay vector of interest, and let $\mathbf{h}=[h_1,\dots,h_Q]^{\top}$ represent the fading coefficients, which are treated as nuisance parameters. Note that the angles are not treated as nuisance parameters, since they are assumed to be deterministically known. According to \cite{nie2024uplink}, the MSE of any unbiased estimator $\hat{\mathbf{l}}$ satisfies $\mathbb{E}\left[(\hat{\mathbf{l}}-\mathbf{l})(\hat{\mathbf{l}}-\mathbf{l})^{\top}\right]\succeq \mathbf{J}^{-1}$,
where $\mathbf{J}\in\mathbb{R}^{Q\times Q}$ is the Fisher information matrix (FIM), whose $(i,j)$-th entry is given by $\mathbf{J}(i,j)
    = \frac{2}{\sigma^2_{\rm s}} \mathrm{Re} \left\{ \mathbf{x}^{\mathrm{H}} 
    \frac{\partial \mathbf{H}_{\rm s}^{\mathrm{H}}}{\partial l_i}
    \frac{\partial \mathbf{H}_{\rm s}}{\partial l_j}
    \mathbf{x} \right\}.$
Taking the derivative yields
\begin{align}
    \mathbf{J}(i,j)
    &= \frac{2}{\sigma^2_{\rm s}} \mathrm{Re} \Bigg\{ 
    h_i^{*} h_j \frac{4\pi^2}{M^2}
    \sum_{m=0}^{M-1} m^2 e^{j2\pi m \frac{l_i-l_j}{M}} \notag \\
    &\qquad\qquad \times 
    \mathbf{x}_m^{\rm H} \mathbf{A}^{\rm H}(\theta_i,\theta_i)
    \mathbf{A}(\theta_j,\theta_j)\mathbf{x}_m
    \Bigg\}.
\end{align}

Since $\mathbf{h}$ is random nuisance parameter, a lower bound can then be partially obtained only for the parameters of interest by using an modified CRB (MCRB), which is defined as $\mathbb{E}\left[(\hat{\mathbf{l}}-\mathbf{l})(\hat{\mathbf{l}}-\mathbf{l})^{\top}\right]\succeq  \tilde{\mathbf{J}}^{-1}$, where $\tilde{\mathbf{J}} = \mathbb{E}_{\mathbf{h}}[\mathbf{J}]$ is the expected modified delay FIM over the nuisance parameters \cite{d2002modified}.
After some manipulation, it can be noticed that the modified FIM $\tilde{\mathbf{J}}$ is a diagonal matrix, due to independence and the zero-mean property of the path coefficients $h_q$. The diagonal entries are given by
\begin{align}
    &\tilde{\mathbf{J}}(i,i)= \notag\\
    &\frac{2}{\sigma^2_{\rm s}} {\rm{Re}}\Bigg\{  \frac{4\pi^2}{M^2Q} \sum_{m=0}^{M-1} m^2 \mathbf{x}_{m}^{\rm H} \mathbf{A}^{\rm H}(\theta_{i},\theta_{i}) \mathbf{A}(\theta_{i},\theta_{i})\mathbf{x}_{m}\Bigg\}. \label{modified_FIM}
\end{align}
During training, rather than directly minimizing the MCRBs, we minimize the logarithmic form of its average value across all sensing targets:
\begin{align}
    \mathcal{L}_{\mathrm{s}}=\log\left(\frac{1}{Q}\mathrm{tr}\left(\tilde{\mathbf{J}}^{-1}\right)\right),
\end{align}
Since the logarithm is strictly increasing, this transformation preserves the desired optimization direction while compressing the dynamic range of the loss. This prevents samples with exceptionally large MCRB values from dominating the training process, thereby improving training stability. The overall training loss is then formulated as
\begin{align}
    \mathcal{L}=(1-\lambda) \mathcal{L}_{\rm c} + \lambda \mathcal{L}_{\rm s}, \label{overall_loss}
\end{align}
where $\lambda \in [0,1]$ controls the trade-off between communication and sensing performance.

\subsection{Training Procedure}
We adopt a training strategy similar to \cite{bian2025lisac}, where the transmitter encoder and communication decoder are trained in an alternating manner. Specifically, the encoder is updated by minimizing the overall loss function $\mathcal{L}$ in \eqref{overall_loss} while keeping the decoder fixed. In turn, the decoder is trained using only the communication loss $\mathcal{L}_{\rm c}$ with the encoder fixed. Moreover, each epoch of training includes more decoder training cycles than encoder training cycles to mitigate decoder performance lag, thereby reducing the risk of convergence to poor local optima. The overall training algorithm is summarized in Algorithm \ref{algo1}.
\begin{algorithm}
\caption{Training algorithm}\label{algo1}
\For{\rm Epochs}{
    \textbf{Fix communication decoder}\\
    \For{\rm Batches}{
        \textbf{Obtain:} $\mathbf{c}, \mathbf{H}_{\rm c}, \boldsymbol{\theta}$;\\
        \textbf{Projection and embeddings:} \\
        \quad bit$\_$token $\in\mathbb{R}^{B,M,N_{\rm h}}$; \\
        \quad channel$\_$token $\in\mathbb{R}^{B,M,N_{\rm h}}$; \\
        \quad angle$\_$token $\in\mathbb{R}^{B,Q,N_{\rm h}}$;\\
        \textbf{Feeding tokens into Transformer:} \\
        \quad encoded$\_$tokens $\in\mathbb{R}^{B,2M+Q,N_{\rm h}}$\\
        \textbf{Feeding first $M$ tokens into MLP:} \\
        \quad $\mathbf{x}\in \mathbb{C}^{B,M,N_{\rm t}}$\\
        \textbf{Calculate $\frac{1}{B}\sum_{b=1}^{B}\mathcal{L}^{(b)}$ and optimize}
    }
    \textbf{Fix transmitter encoder}\\
    \For{\rm Batches}{
    \textbf{Construct:} $\mathbf{u}\in\mathbb{R}^{B,M,2N_{\rm r}N_{\rm t}+2N_{\rm t}+2N_{\rm t}^{2}+1}$; \\
    \textbf{Projection and embeddings:} token $\in\mathbb{R}^{B,M,N_{\rm h}}$; \\
    \textbf{Feeding tokens into Transformer and MLP:} \\
    \quad soft bit probabilities $\mathbf{p}\in(0,1)^{B\times M\times C}$; \\
    \textbf{Calculate $\frac{1}{B}\sum_{b=1}^{B}\mathcal{L}^{(b)}_{\rm c}$ and optimize}
    }
}
\end{algorithm}

\section{Numerical Results}



Unless otherwise specified, we use $C=6$, $M=12$, $N_{\rm t}=N_{\rm r}=9$, $P=5$, and $Q=3$. The normalized communication-path delays are independently and uniformly distributed over $[0,5]$. The angles are drawn independently from the grid $\{-60^{\circ},-57^{\circ},\dots,60^{\circ}\}$. The transmitter and receiver each use a Transformer with hidden dimension $N_{\rm h}=128$, four attention heads, four layers, and zero dropout. Training uses a batch size of $B=256$ for $240$ epochs. Each epoch consists of one transmitter training cycle followed by six receiver training cycles. Each cycle uses $3\times10^4$ samples, each containing $MC=72$ information bits. The Adam optimizer is used with cosine annealing of the learning rate from $10^{-4}$ to $5\times10^{-6}$. The communication loss in \eqref{overall_loss} is scaled by $10$, giving the transmitter training objective $\mathcal{L}=10(1-\lambda)\mathcal{L}_{\rm c}+\lambda\mathcal{L}_{\rm s}$.
Separate models are trained for $\lambda\in\{0,0.2,0.4,0.6,0.8,1\}$. For each training codeword, the communication SNR is independently sampled uniformly over $[10,26]$~dB. During transmitter updates, the sensing SNR is independently sampled with equal probability from $\{-15,-10,-5,0,5,10\}$~dB. We define the nominal SNR as ${\rm SNR}=E_{\rm s}/\sigma^2$, where $E_{\rm s}=1$ is the average transmit energy per antenna and $\|\mathbf{x}_m\|^2=N_{\rm t}=9$ on each subcarrier.

\subsection{Baselines}
We compare the proposed scheme with convolutionally coded QPSK (CC-QPSK) and 8-PSK (CC-8PSK). Both benchmarks adopt SVD-based precoding, which is widely used in MIMO-OFDM systems with CSIT. To avoid allocating power to extremely weak eigenmodes, CC-QPSK and CC-8PSK use the five and four strongest singular modes per subcarrier, respectively. Both employ an interleaved convolutional code with constraint length $7$ and octal generators $(171,133)$, terminated with six zero-tail bits and punctured from rate $1/2$ to a nominal rate of $2/3$. The coded bits are Gray-mapped to unit-energy symbols, with three and $27$ zero-padding bits added for CC-QPSK and CC-8PSK, respectively. Including termination and padding, the effective payload rates are $3/5$ and $1/2$. Soft-decision Viterbi decoding is used at the receiver, and BER is computed over the original $72$ information bits.

\begin{figure}
    \centering
    \includegraphics[width=0.9\linewidth]{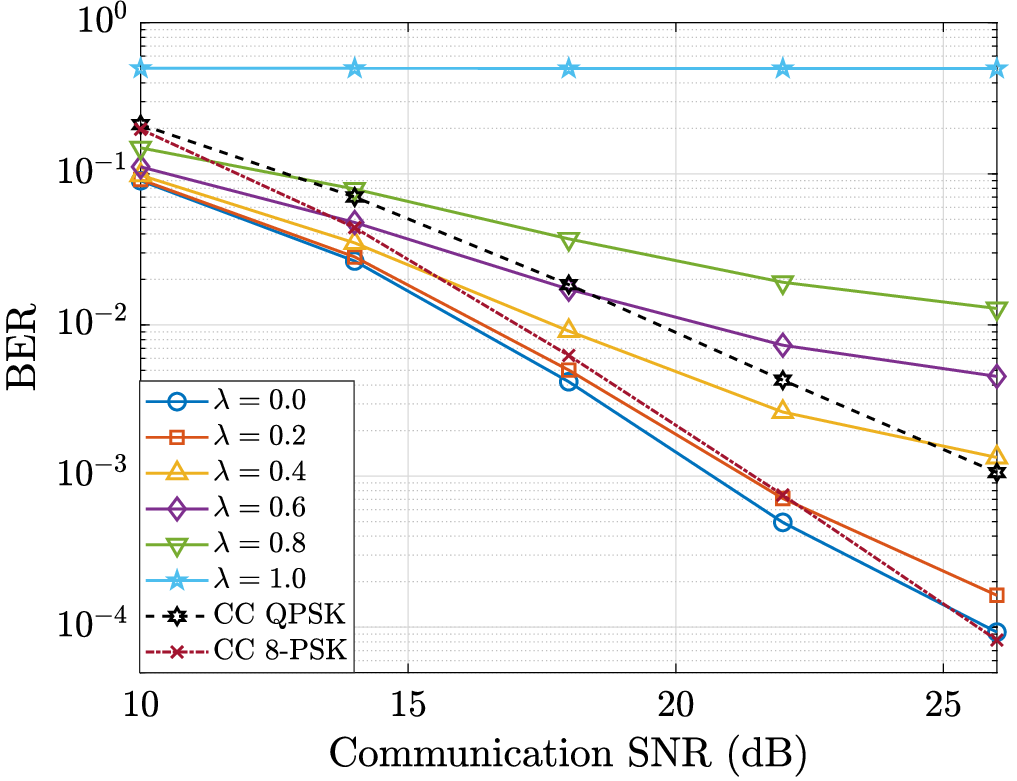}
    \caption{BER comparison between different $\lambda$ and baselines.}
    \label{fig:BER}
\end{figure}

\begin{figure}
    \centering
    \includegraphics[width=0.9\linewidth]{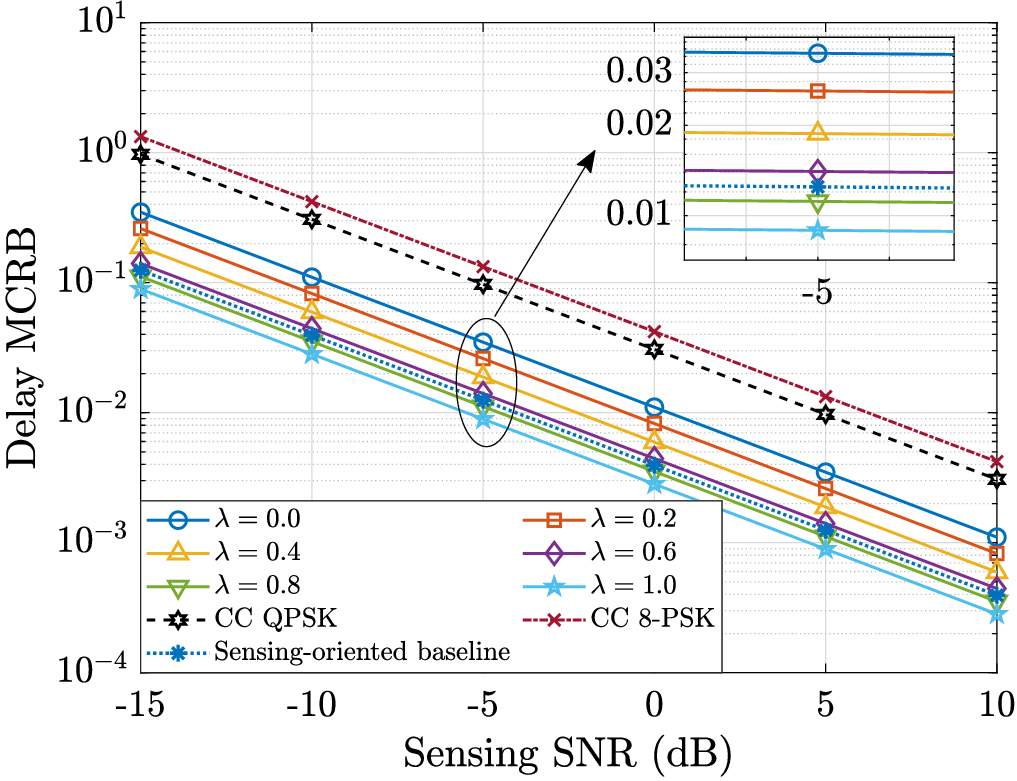}
    \caption{MCRB comparison between different $\lambda$ and baselines.}
    \label{fig:MCRB}
\end{figure}

To further examine the sensing capability of the proposed scheme, we construct a sensing-oriented transmit waveform based on the known sensing angles. It is evident from \eqref{modified_FIM} that the delay FIM is closely related to the term $\sum_{q=1}^{Q}\mathbf{x}_{m}^{\rm H} \mathbf{A}^{\rm H}(\theta_{q},\theta_{q}) \mathbf{A}(\theta_{q},\theta_{q})\mathbf{x}_{m}$, which suggests that the transmit energy should be concentrated in the angular subspace spanned by the target steering vectors. Motivated by this observation, we define the angular covariance matrix as
\begin{align}
    \mathbf{R}_{\theta} &= \sum_{q=1}^{Q}\mathbf{A}^{\rm H}(\theta_{q},\theta_{q}) \mathbf{A}(\theta_{q},\theta_{q})=\sum_{q=1}^{Q}\mathbf{a}^{*}_{\rm t}(\theta_{q}) \mathbf{a}^{\top}_{\rm t}(\theta_{q}).
\end{align}
Then, perform the eigenvalue decomposition $\mathbf{R}_{\theta} = \mathbf{U}_{\theta} \boldsymbol{\Lambda}_{\theta}  \mathbf{U}^{\rm H}_{\theta}$,
where the columns of $\mathbf{U}_{\theta}$ are the eigenvectors $\mathbf{u}_{i}$ of $\mathbf{R}_{\theta}$. Let $\mathbf{U}_{Q}$ contain the $Q$ dominant eigenvectors. For the $m$-th subcarrier, the sensing-oriented waveform is generated as $\mathbf{x}_{m}=\mathbf{U}_{Q}\mathbf{f}_{m}$, 
where $\mathbf{f}_{m}=\sqrt{\frac{N_{\rm t}}{Q}}[1, e^{j2\pi m/M}, \dots, e^{j2\pi m (Q-1)/M}]^{\top}$ represents the beam coefficients applied to the spatial basis $\mathbf{U}_{Q}$ across different subcarriers. This waveform carries no information bits and is evaluated only as a sensing reference.

\subsection{Comparisons}
Fig.~\ref{fig:BER} shows how the communication performance varies with the weighting factor. Specifically, the proposed scheme with $\lambda=0$ outperforms CC-8PSK over most of the evaluated SNR range, while $\lambda=0$, $0.2$, and $0.4$ consistently outperform CC-QPSK. Since CC-8PSK avoids the weaker fifth mode used by CC-QPSK and concentrates power on four stronger modes, CC-8PSK achieves better communication performance than CC-QPSK. For the proposed scheme, increasing $\lambda$ generally degrades BER, since assigning a larger weight to the sensing task will inevitably degrade communication performance. Moreover, Fig.~\ref{fig:MCRB} shows that communication-oriented waveforms generally yield relatively large delay MCRB values. This is expected because SVD-based precoding concentrates energy on communication eigenmodes that do not necessarily align with the sensing-target directions. CC-QPSK nevertheless achieves a lower MCRB than CC-8PSK. Its additional spatial mode may couple effectively to sensing targets even when its communication gain is weak. All learned designs achieve lower MCRB than both communication benchmarks, while the designs with $\lambda=0.8$ and $1$ also outperform the sensing-oriented baseline. In particular, $\lambda=0.4$ provides an attractive balance between communication reliability and sensing performance, achieving BER close to the CC-QPSK baseline and a delay MCRB approaching that of the sensing-oriented baseline. Overall, the results illustrate how the learned codewords realize different BER--MCRB trade-offs through the weighting factor $\lambda$, confirming the effectiveness of the proposed joint waveform design.

\begin{figure*}[!th]	
	\centering   
	    \subfigure[$\lambda=0$]{\includegraphics[scale=0.22]{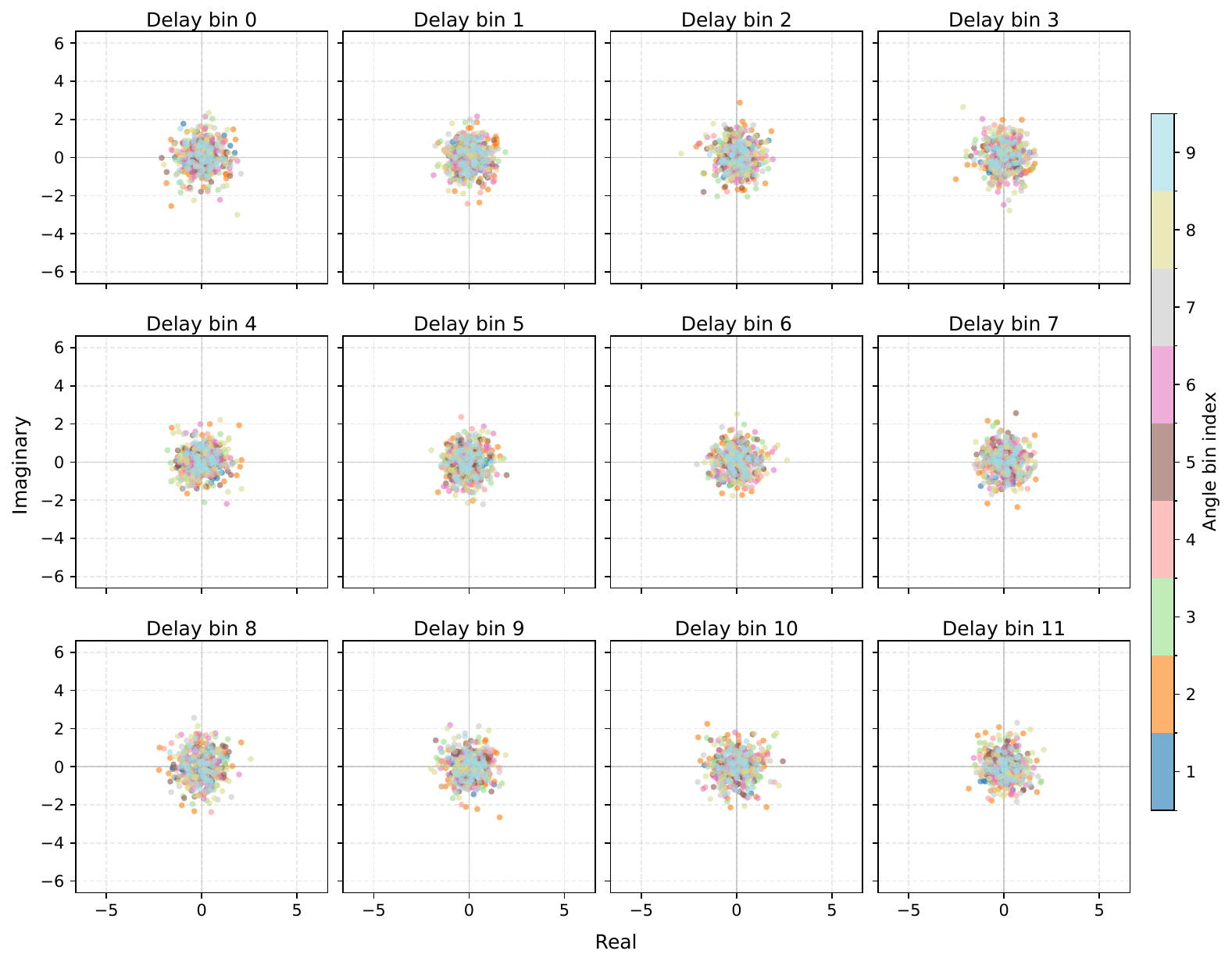}}
    	\subfigure[$\lambda=0.4$]{\includegraphics[scale=0.22]{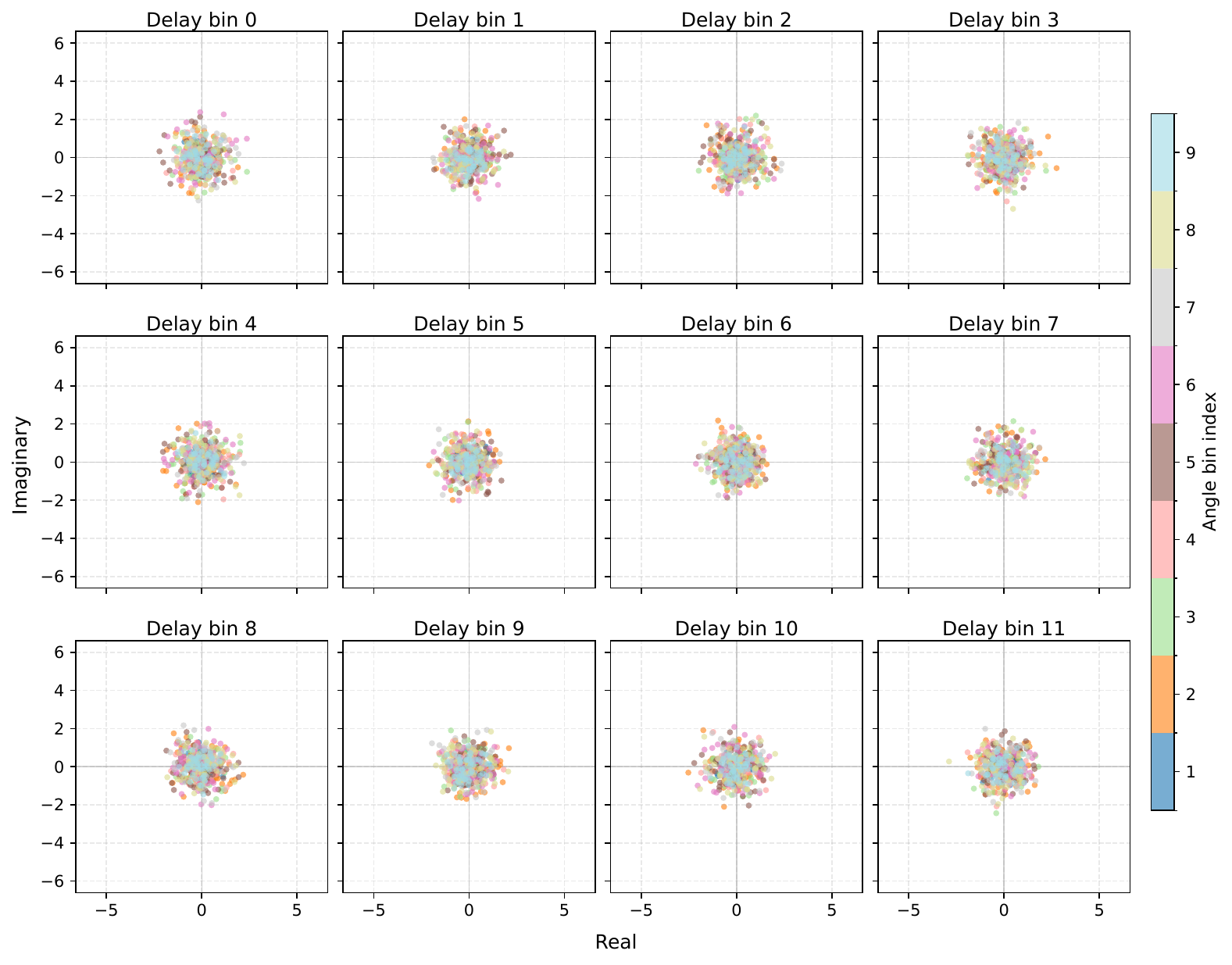}}
        \subfigure[$\lambda=1$]{{\includegraphics[scale=0.22]{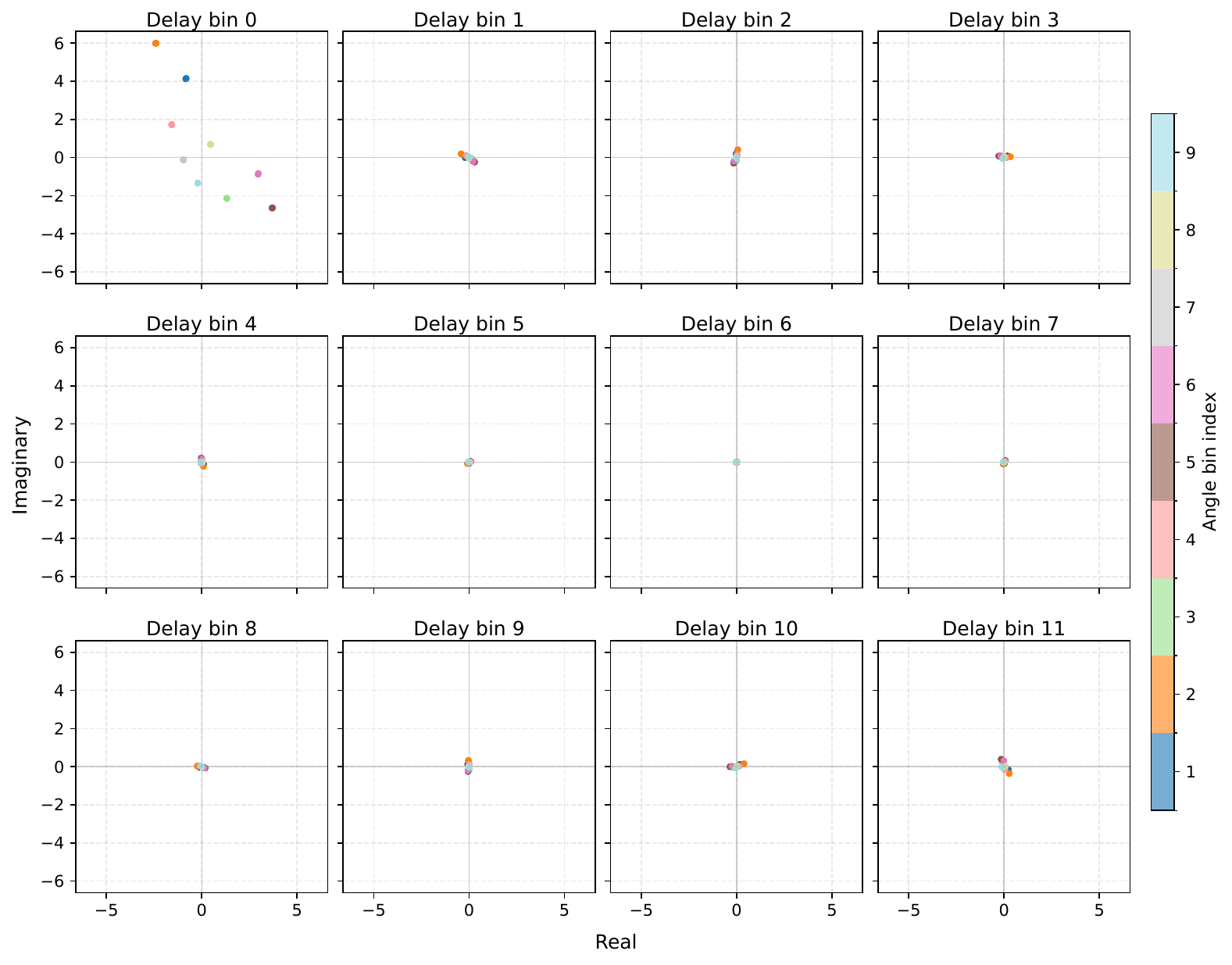}}}
    \caption{Visualization of the learned codewords in the delay domain.} 
    \label{fig:constellation_delays}
\end{figure*} 
\begin{figure*}[!th]	
	\centering   
	    \subfigure[$\lambda=0$]{\includegraphics[scale=0.28]{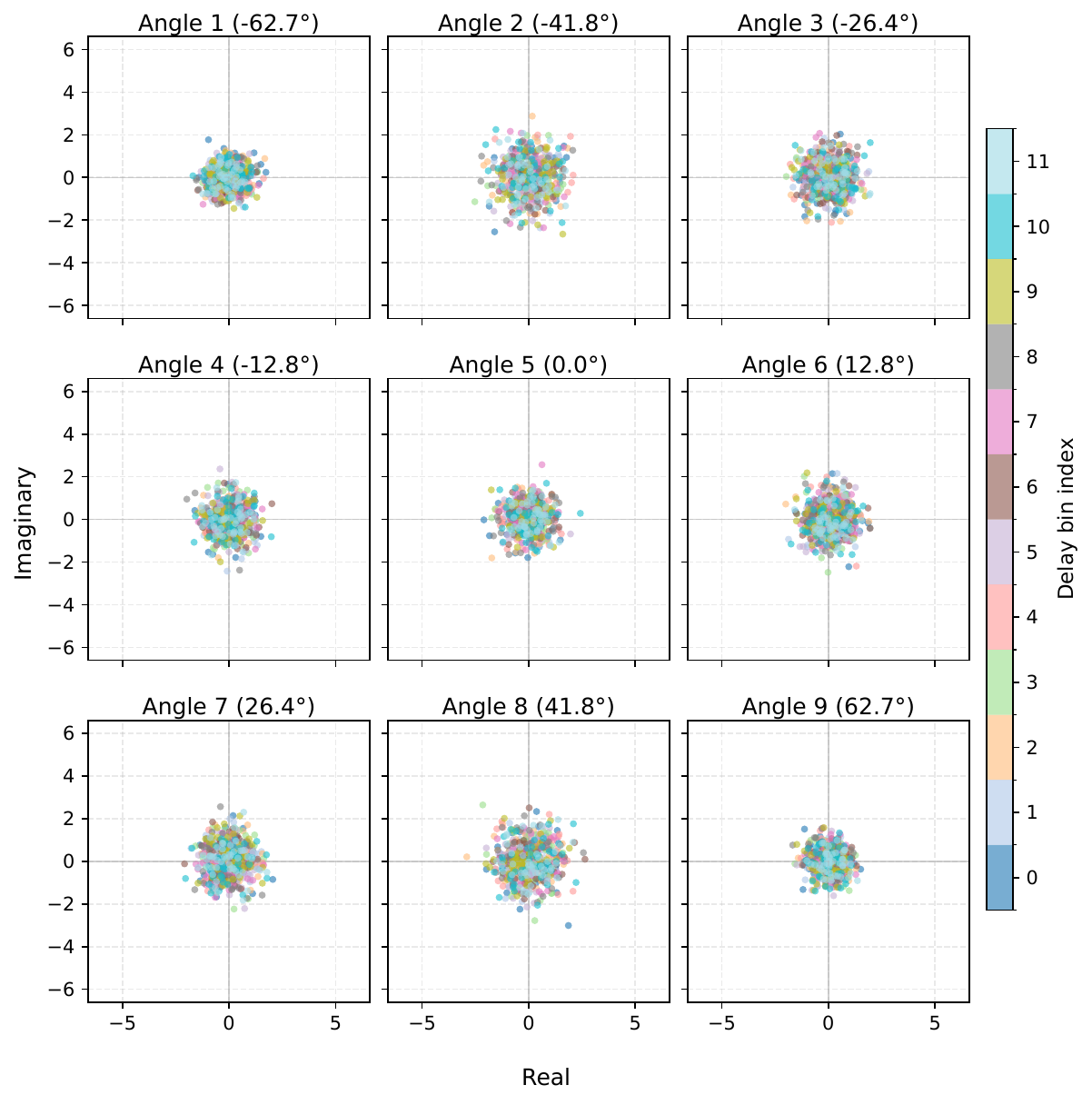}}
    	\subfigure[$\lambda=0.4$]{\includegraphics[scale=0.28]{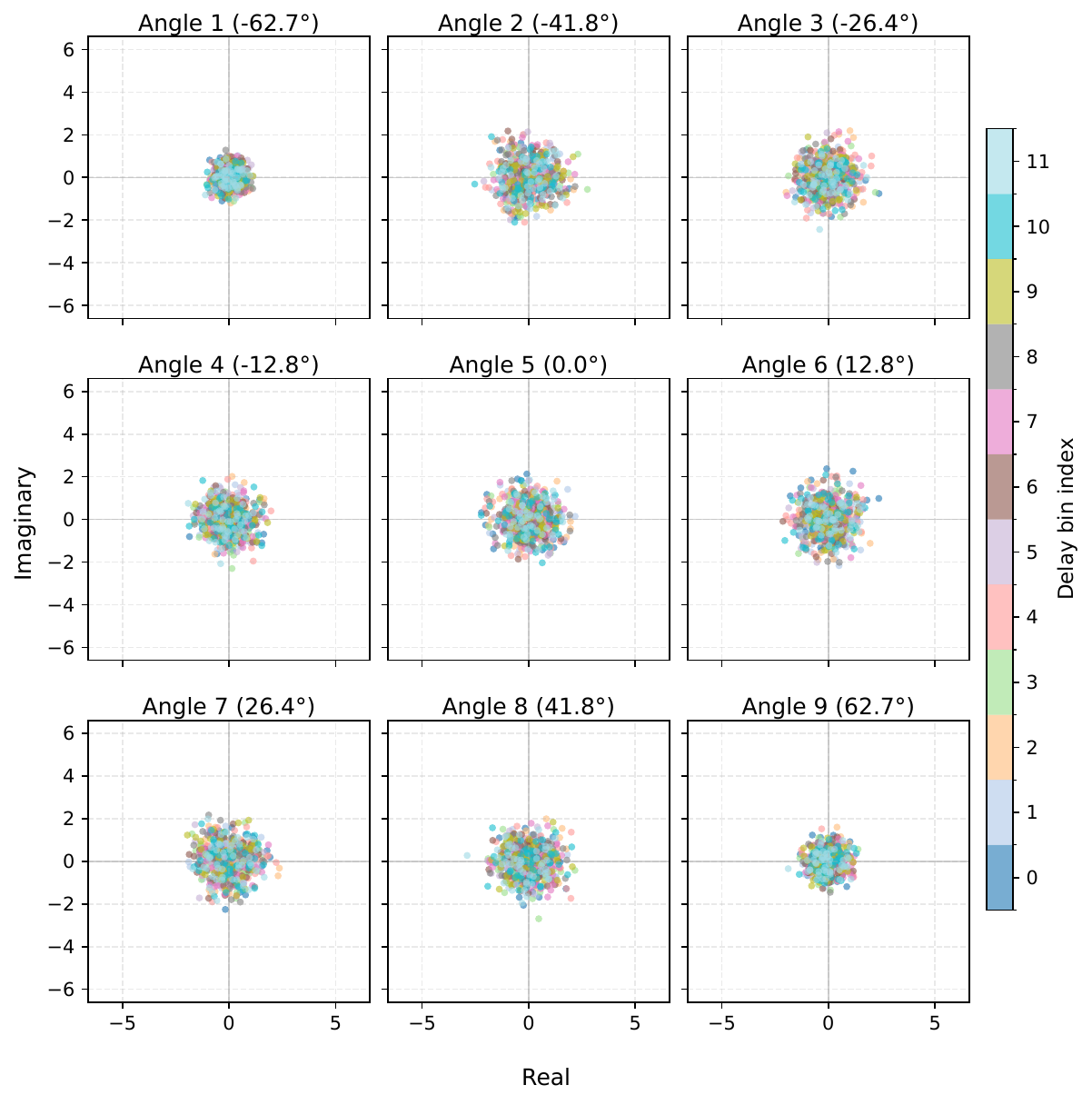}}
        \subfigure[$\lambda=1$]{{\includegraphics[scale=0.28]{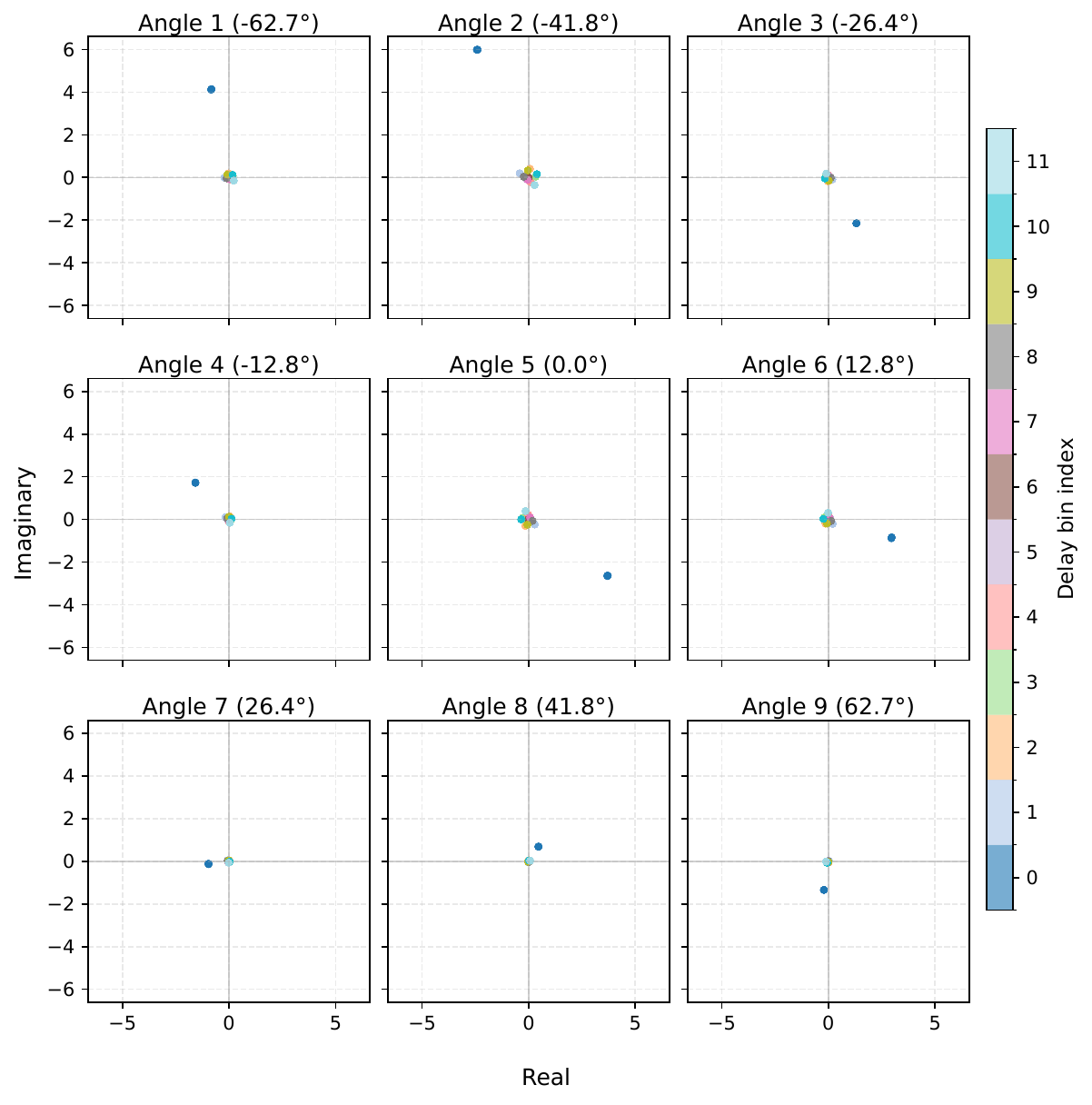}}}
    \caption{Visualization of the learned codewords in the angle domain.} 
    \label{fig:constellation_angles}
\end{figure*} 

\subsection{Analysis and Insights}

To examine the learned waveform structure, we visualize $50$ codewords randomly sampled from the codebook, using the same communication channel and target angles for all values of $\lambda$. For the communication-oriented design ($\lambda=0$), Fig.~\ref{fig:constellation_delays} shows that different codewords yield widely dispersed constellation points within each delay bin. For the sensing-oriented design ($\lambda=1$), energy concentrates predominantly in the first delay bin, producing an impulse-like waveform that can be advantageous for delay estimation. Its coefficients remain nearly unchanged across codewords, as the sensing-only objective does not require large distances between codewords to ensure reliable message discrimination. Fig.~\ref{fig:constellation_angles} illustrates the angular power distribution for targets at $-51^{\circ}$, $-48^{\circ}$, and $6^{\circ}$. The sensing-oriented design ($\lambda=1$) concentrates power in angular bins near the target directions, whereas the communication-oriented design ($\lambda=0$) distributes power more evenly across the bins. The $\lambda=0.4$ design represents an intermediate case, with its angular power distribution shaped by both communication and sensing objectives.

\section{Conclusion}
In this paper, we explored code designs for ISAC transmissions in MIMO-OFDM systems, where an attractive trade-off between communication and sensing performance is achieved through the proposed DL-based framework. In particular, the designed NN efficiently handles heterogeneous and high-dimensional inputs to generate transmit ISAC  waveforms. By adjusting the weighting factor in the joint loss function, different models are trained to realize various communication and sensing trade-offs. The communication-oriented design produces codewords with energy distributed across delay bins, whereas the sensing-oriented design forms impulse-like waveforms with power concentrated near the target directions. The balanced ISAC waveform naturally exhibits an intermediate structure between these two extremes. Numerical results demonstrated that the learned coded waveforms significantly outperform conventional CC-QPSK baselines in terms of MCRB performance, without compromising BER performance. Moreover, an appealing operating point was identified, where the proposed scheme achieves performance close to both the communication-oriented and sensing-oriented baselines simultaneously.


\bibliographystyle{IEEEtran}
\bibliography{ref}

@IEEEtranBSTCTL{bibfbcs:BSTcontrol,
	CTLuse_forced_etal = "no",
	CTLmax_names_forced_etal = "1",
        CTLnames_show_etal       = "1"
}

@article{nie2024uplink,
  title={Uplink multi-user {OTFS}: Transmitter design based on statistical channel information},
  author={Nie, Mingcheng and Li, Shuangyang and Mishra, Deepak and Yuan, Jinhong and Ng, Derrick Wing Kwan},
  journal={IEEE Trans. Commun.},
  volume={73},
  number={7},
  pages={4678--4696},
  month=jul,
  year={2025},
}

@article{d2002modified,
  title={The modified {Cramer-Rao} bound and its application to synchronization problems},
  author={D'Andrea, Aldo N and Mengali, Umberto and Reggiannini, Ruggero},
  journal={IEEE Trans. Commun.},
  volume={42},
  number={2/3/4},
  pages={1391--1399},
  year={1994},
  month={Feb.--Apr.},
}

@inproceedings{bian2025lisac,
  title={{LISAC}: Learned coded waveform design for {ISAC} with {OFDM}},
  author={Bian, Chenghong and Zhang, Yumeng and G{\"u}nd{\"u}z, Deniz},
  booktitle={Proc. IEEE WCNC},
  pages={1--7},
  year={2025},
}

@article{vaswani2017attention,
  title={Attention is all you need},
  author={Vaswani, Ashish and Shazeer, Noam and Parmar, Niki and Uszkoreit, Jakob and Jones, Llion and Gomez, Aidan N and Kaiser, {\L}ukasz and Polosukhin, Illia},
  journal={Proc. NeurIPS},
  volume={30},
  pages={5998--6008},
  year={Dec. 2017}
}

@inproceedings{devlin2019bert,
  title={{BERT}: Pre-training of deep bidirectional transformers for language understanding},
  author={Devlin, Jacob and Chang, Ming-Wei and Lee, Kenton and Toutanova, Kristina},
  booktitle={Proc. NAACL-HLT},
  pages={4171--4186},
  year={2019}
}

@inproceedings{kim2024short,
  title={Short-length code designs for integrated sensing and communications using deep learning},
  author={Kim, Muah and Jahani-Nezhad, Tayyebeh and Li, Shuangyang and Schaefer, Rafael F and Caire, Giuseppe},
  booktitle={Proc. IEEE ICC},
  pages={3536--3541},
  year={2024},
}

@ARTICLE{nie11373535standard,
  author={Nie, Mingcheng and Chong, Ruoxi and Li, Shuangyang and Farhang, Arman and Göttsch, Fabian and Ng, Derrick Wing Kwan and Matthaiou, Michail and Li, Yonghui},
  journal={IEEE Commun. Stand. Mag.}, 
  title={Toward Standardizing {OTFS}: A Candidate Waveform for Next-Generation Wireless Networks}, 
  month=jun,
  year={2026},
  volume={10},
  number={2},
  pages={107--118},
  doi={10.1109/MCOMSTD.2026.3657606}}

@article{liu2022integrated,
  title={Integrated sensing and communications: Toward dual-functional wireless networks for {6G} and beyond},
  author={Liu, Fan and Cui, Yuanhao and Masouros, Christos and Xu, Jie and Han, Tony Xiao and Eldar, Yonina C and Buzzi, Stefano},
  journal={IEEE J. Select. Areas Commun.},
  volume={40},
  number={6},
  pages={1728--1767},
  year={Jun. 2022},
}

@article{xiong2023fundamental,
  title={On the fundamental tradeoff of integrated sensing and communications under {Gaussian} channels},
  author={Xiong, Yifeng and Liu, Fan and Cui, Yuanhao and Yuan, Weijie and Han, Tony Xiao and Caire, Giuseppe},
  journal={IEEE Trans. Info. Theory},
  volume={69},
  number={9},
  pages={5723--5751},
  year={Sept. 2023},
  publisher={IEEE}
}

@article{aditya2025channel,
  title={Channel coding meets sequence design via machine learning for integrated sensing and communications},
  author={Aditya, Sundar and Varasteh, Morteza and Clerckx, Bruno},
  journal={arXiv preprint arXiv:2503.23119},
  year={2025}
}

@inproceedings{nie2025novel,
  title={A novel cross-domain channel estimation scheme for {OFDM}},
  author={Nie, Mingcheng and Chong, Ruoxi and Li, Shuangyang and Yuan, Weijie and Ng, Derrick Wing Kwan and Matthaiou, Michalis and Caire, Giuseppe and Li, Yonghui},
  booktitle={Proc. IEEE GLOBECOM},
  pages={5133-5138},
  year={2025},
}

@article{liu2018toward,
  title={Toward dual-functional radar-communication systems: Optimal waveform design},
  author={Liu, Fan and Zhou, Longfei and Masouros, Christos and Li, Ang and Luo, Wu and Petropulu, Athina},
  journal={IEEE Trans. Signal Process.},
  volume={66},
  number={16},
  pages={4264--4279},
  year={Aug. 2018},
}

@inproceedings{nie2026neural,
  author    = {Mingcheng Nie and Hao Chang and Shuangyang Li
               and Haiyao Yu and Jiafu Hao and Yonghui Li},
  title     = {Neural network-based delay-{Doppler}-assisted
               channel estimation for {OFDM}},
  booktitle = {Proc. Int. Symp. Wireless Commun. Syst. (ISWCS)},
  year      = {2026},
  note      = {to appear}
}

@inproceedings{nie2026refinement,
  author    = {Mingcheng Nie and Hao Chang and Xiaoqi Zhang
               and Junkai Liu and Wibowo Hardjawana
               and Branka Vucetic and Yonghui Li},
  title     = {Channel estimation for {OFDM} via
               delay--{Doppler} refinement},
  booktitle = {Proc. IEEE Int. Workshop Signal Process.
               Artif. Intell. Wireless Commun. (SPAWC)},
  year      = {2026},
  note      = {to appear}
}



\end{document}